\documentclass[aps,showpacs,twocolumn,superscriptaddress]{revtex4}
\usepackage{amssymb}
\usepackage{color}
\usepackage{graphicx,amsmath}
\usepackage[normalem]{ulem}
\usepackage{color,bm}
\usepackage{graphicx,amsmath}
\usepackage[colorlinks]{hyperref}
\hypersetup{colorlinks,citecolor=red,linkcolor=blue,urlcolor=blue}

\begin{document}
\title{Thermodynamic, dynamic and transport properties of quantum spin liquid
in herbertsmithite from experimental and theoretical point of
view}
\author{V. R. Shaginyan}\email{vrshag@thd.pnpi.spb.ru}
\affiliation{Petersburg
Nuclear Physics Institute of NRC "Kurchatov Institute",
Gatchina, 188300, Russia}\affiliation{Clark Atlanta University,
Atlanta, GA 30314, USA} \author{A. Z. Msezane}\affiliation{Clark
Atlanta University, Atlanta, GA 30314, USA} \author{M. Ya.
Amusia}\affiliation{Racah Institute of Physics, Hebrew
University, Jerusalem 91904, Israel}\affiliation{Ioffe Physical
Technical Institute, RAS, St. Petersburg 194021, Russia}
\author{J.~W.~Clark}
\affiliation{McDonnell Center for the Space Sciences \&
Department of Physics, Washington University, St.~Louis, MO
63130, USA}
\affiliation{
Centro de Investiga\c{c}\~{a}o em Matem\'{a}tica e Aplica\c{c}\~{o}es,
University of Madeira, 9020-105 Funchal, Madeira, Portugal}
\author{G. S. Japaridze}\affiliation{Clark Atlanta
University, Atlanta, GA 30314, USA}\author{V. A.
Stephanovich}\affiliation{Institute of Physics, Opole
University,\\Oleska 48, 45-052, Opole, Poland}
\author{Y. S. Leevik} \affiliation{National Research University Higher School
of Economics, St.Petersburg, 194100, Russia}

\begin{abstract}

In our review we focus on the quantum spin liquid (QSL),
defining the thermodynamic, transport and relaxation properties
of geometrically frustrated magnet (insulators) represented by
herbertsmithite $\rm ZnCu_{3}(OH)_6Cl_2$. QSL is a quantum state
of matter having neither magnetic order nor gapped excitations
even at zero temperature. QSL along with heavy fermion metals
can form a new state of matter induced by the topological
fermion condensation quantum phase transition. The observation
of QSL in actual materials such as herbertsmithite is of
fundamental significance both theoretically and technologically,
as it could open a path to creation of topologically protected
states for quantum information processing and quantum
computation. It is therefore of great importance to establish
the presence of a gapless QSL state in one of the most
prospective material herbertsmithite. In this respect,
interpretation of current theoretical and experimental studies
of herbertsmithite are controversial in their implications.
Based on published experimental data augmented by our
theoretical analysis, we present evidence for the the existence
of a QSL in the geometrically frustrated insulator
herbertsmithite $\rm ZnCu_{3}(OH)_6Cl_2$, providing a strategy
for unambiguous identification of such a state in other
materials. To clarify the nature of QSL in herbertsmithite, we
recommend measurements of heat transport, low-energy inelastic
neutron scattering, and optical conductivity $\overline{\sigma}$
in $\rm ZnCu_{3}(OH)_6Cl_2$ crystals subject to an external
magnetic field at low temperatures. Our analysis of the behavior
of $\overline{\sigma}$ in herbertsmithite justifies this set of
measurements, which can provide conclusive experimental
demonstration of the nature of its spinon-composed quantum spin
liquid. Theoretical study of the optical conductivity of
herbertsmithite allows us to expose the physical mechanisms
responsible for its temperature and magnetic-field dependence.
We also suggest that artificially or spontaneous introducing
inhomogeneity at nanoscale into $\rm ZnCu_{3}(OH)_6Cl_2$ can
both stabilize its QSL and simplify its chemical preparation,
and can provide for tests that elucidate the role of impurities.
We make predictions of the results of specified measurements
related to the dynamical, thermodynamic and transport properties
in the case of a gapless QSL.
\end{abstract}

\pacs{64.70.Tg, 75.40.Gb, 78.20.-e, 71.10.Hf\\
Keywords: quantum spin liquids; herbertsmithite; fermion
condensation; topological quantum phase transitions; flat bands}

\maketitle

\tableofcontents

\section{Introduction}

The frustrated magnet (insulator) herbertsmithite $\rm
ZnCu_3(OH)_6Cl_2$ is one of the best candidates for
identification as a material that hosts a quantum spin liquid
(QSL), thereby determining the nature of its thermodynamic,
relaxation and transport properties. The insulating nature of
$\rm ZnCu_3(OH)_6Cl_2$ has been established: There is a 3.3 eV
charge gap, see e.g. \cite{pustogow2017,pustogow2018}. In our
review we focus on a spin gap of QSL, and for brevity will call
it "spin gap". At low temperatures $T$, a QSL may have or may
not have a gap in its excitation spectrum of spinons, which are
fermion quasiparticles of zero charge that occupy the
corresponding Fermi sphere with Fermi momentum $p_F$. We note
that, in contrast to metals, spinons cannot support charge
current but these, for instance, can carry heat, as electrons of
metals do. The influence of a gap on the properties is huge, for
at low $T$ the properties of herbertsmithite are similar to
those of common insulators, while the properties resemble those
of metals, provided that the gap is absent. Thus, it is a
challenge to experimentally establish the presence of the gap
and its value, for current theoretical and experimental studies
of herbertsmithite are controversial, and cannot give definite
answer to the challenge. In our review we offer a number of
experiments that can allow one to unambiguously test the
presence of gapless QSL.

In a magnet with geometric frustration, where the spins cannot
be ordered even at temperatures close to absolute zero, they are
staying in a liquid quantum spin state. The herbertsmithite is
an antiferromagnet with a kagome lattice of spins $S = 1/2$.
Recent experimental studies have shown his unusual properties
\cite{helt,herb2,herb3,herb,t_han:2012,t_han:2014}.

The balance of electrostatic forces for the $\rm Cu^{2+}$ ions
in the kagome structure is such that they occupy distorted
octahedral sites. The magnetic planes formed by the $\rm
Cu^{2+}$ $S=1/2$ ions are interspersed with nonmagnetic $\rm
Zn^{2+}$ layers. In samples, $\rm Cu^{2+}$ defects occupy the
nonmagnetic $\rm Zn^{2+}$ sites between the kagome layers with
$x\simeq 15\%$ probability, thus introducing randomness and
inhomogeneity into the lattice \cite{Han}. As we shall see, the
starring role of impurities in formation of QSL is not clearly
understood, and now is under thorough investigation
\cite{Pad2017,Sait2018,Lazic2018,Pint2018,Shi2019,screp-17}.
However, we suggest that the influence of impurities induced by
the homogeneity on the properties of $\rm ZnCu_3(OH)_6Cl_2$ can
be tested by varying $x$. We note that the impurities randomly
located at nanoscale level can support QSL as it is observed in
measurements on $\rm Zn_xCu_{4-x}OH_6Cl_2$ with $1\geq
x\geq0.8$, while at $x>0.8$ glassy dynamics emerges
\cite{herb2}. Moreover, the impurities randomly located at
nanoscale level can stabilize QSL as it is observed in
measurements on the verdazyl-based complex $\rm
Zn(hfac)_2(A_xB_{1-x})$ \cite{screp-17}, and make QSL stable  in
high magnetic fields, as it takes place in case  $\rm Mg$-doped
$\rm SrCu_2(BO_3)_2$ \cite{Shi2019}.

The experiments made on $\rm ZnCu_3(OH)_6Cl_2$ have not found
any traces of magnetic order in it. Nor they have found the spin
freezing down to temperatures of around 50 mK. In these
respects, herbertsmithite is the best candidate among quantum
magnets to contain QSL described above
\cite{helt,herb2,herb3,herb,t_han:2012,t_han:2014}. These
assessments are supported by model calculations indicating that
an antiferromagnet on kagome lattice has gapless spin liquid
ground state
\cite{prr,shaginyan:2011,shaginyan:2012:A,shaginyan:2011:C,shaginyan:2013:D,book,hfliq,Normand}.
At the same time, recent suggestion \cite{Han,Han11,sc_han},
that there can exist a small spin-gap in the kagome layers may
stand in conflict to this emerging picture (see also
Refs.~\cite{norman,zhou,savary} for a recent review). The
obtained results is a combination of experimental and
theoretical ones. Latter have been obtained in the framework of
the model, which takes the Cu impurities presence into account.
The experimental data has been obtained by  inelastic neutron
scattering on $\rm ZnCu_{3}(OH)_6Cl_2$ crystals. It is assumed
that the influence of the Cu impurity ensemble on the observed
properties of herbertsmithite may be disentangled from that of
the kagome lattice geometry \cite{Han,Han11,sc_han}. It is
further assumed that the impurity ensemble may be represented as
a result of dilution of some prototypic cubic lattice. The model
supposes that the spin gap occurs in magnetic fields up to 9 T.
We note that measurements of the local spin susceptibility of
$\rm ZnCu_{3}(OH)_6Cl_2$ show that the kagome spins exhibit a
spin gap \cite{sc_han}. While the measurements of the bulk spin
susceptibility evidences the absence of the gap \cite{herb3}. We
shall see below that it is not possible to separate
contributions coming from the local and bulk susceptibility, for
both the kagome and the impurities spins form an integral
system.

At the same time, without magnetic field, the spin
susceptibility $\chi$ shows the behavior, in many respects
similar to the Curie law. Latter behavior demonstrates that the
copper spin ensemble plays a role of weakly interacting
impurities \cite{Han,Han11,sc_han}. The same behavior is
recently reported in a new kagome quantum spin liquid candidate
$\rm Cu_3Zn(OH)_6FBr$ \cite{feng}. As a result, we observe a
challenging contradiction between two sets of experimental data
when some of them state the absent of a gap, while the other
present evidences in the favor of gap. Subsequently we shall
demonstrate that the model based on Cu spin ensemble, is rather
synthetic. In other words, this model does not have a
possibility to distinguish the Cu ensemble and kagome lattice
contributions. {\it This is because the impurities, being
embedded in kagome host lattice, form a single integral entity
at nanoscale.}  The above model may contradict the accumulated
knowledge about the physical properties of $\rm
ZnCu_{3}(OH)_6Cl_2$.  This knowledge is the result of vast
theoretical and experimental efforts related to static and
dynamic properties of herbertsmithite
\cite{helt,herb2,herb3,herb,t_han:2012,t_han:2014,prr,shaginyan:2011,shaginyan:2012:A,shaginyan:2011:C,shaginyan:2013:D,book,hfliq,Normand}.
Thus, when analyzing the physics of quantum spin liquid in the
herbertsmithite, it is of crucial importance to verify the
existence of a spin gap by experimental and theoretical studies,
for the gap strongly influences all its thermodynamic, transport
and relaxation properties. To analyze QSL behavior
theoretically, we employ the strongly correlated quantum spin
liquid (SCQSL)
\cite{shaginyan:2011,hfliq,prr,shaginyan:2011:C,book} model. A
simple kagome lattice may host a dispersionless topologically
protected branch of the quasiparticle spectrum with zero
excitation energy, known as a flat band
\cite{ks91,khod1994,shaginyan:2011,hfliq,prr,volovik}. In that
case the topological fermion condensation quantum phase
transition (FCQPT) can be considered as a quantum critical point
(QCP) of the $\rm ZnCu_3(OH)_6Cl_2$ spinon-composed QSL. Spinons
have zero charge, occupying the Fermi sphere up to the Fermi
momentum $p_F$ \cite{prr,book,volovik}. Taking into account that
we are dealing with the real chemical compound $\rm
ZnCu_3(OH)_6Cl_2$ rather than with an ideal kagome lattice, we
have to bear in mind that the actual magnetic interaction in the
compound can shift QSL away from FCQPT, before or beyond QCP.
Thus, the actual critical point location has to be established
by experimental data analysis. The real part of the optical
conductivity $\overline{\sigma}$, measured at low frequency on
geometrically frustrated magnetic insulators, can yield
important experimental clues to the nature of spinon-based QSL
\cite{Pilon,Lee}, especially at wide range of temperatures $T$
and magnetic fields $B$. Note that the consistent interpretation
of the above data is a difficult theoretical problem
\cite{Pilon}.

The main aim of the present review is to expose QSL as a new
state of matter, also formed by heavy fermion (HF) metals, and
to attract attention to experimental studies of $\rm
ZnCu_3(OH)_6Cl_2$ that have the potential of revealing both the
underlying physics of QSL and the presence or absence of a gap
in spinon excitations that form the thermodynamic, transport and
relaxation properties. To clarify the nature of QSL in
herbertsmithite, we recommend measurements of heat transport,
low-energy inelastic neutron scattering, and optical
conductivity $\overline{\sigma}$ in $\rm ZnCu_{3}(OH)_6Cl_2$
single crystals subject to an external magnetic field at low
temperatures. We show that QSL is situated near FCQPT, which
stems dependence of $\overline{\sigma}$ on the external magnetic
field. We argue that the general physical picture inherent in
QSL can be "purified'' of the microscopic contributions coming
both from phonons and the impurities polluting each specific
sample being studied. We suggest that the influence of
impurities, induced by the inhomogeneity, on the properties of
$\rm ZnCu_3(OH)_6Cl_2$ can be tested by varying $x$.

In the following sections on thermodynamic properties,
relaxation and optical conductivity, we shall demonstrate that
the existence of a gap within the impurity model of
herbertsmithite may contradict recent experimental data
collected on this material; instead, the impurities and kagome
planes can form a genuine SCQSL. Accordingly, the observed gap
\cite{sc_han} related to the kagome planes may not be a real one
as it is not a physical mechanism for the observed
thermodynamic, relaxation and conductivity properties of $\rm
ZnCu_3(OH)_6Cl_2$.

\section{Frustrated insulator Herbertsmithite} \label{HB2}

\begin{figure} [! ht]
\begin{center}
\includegraphics [width=0.55\textwidth]{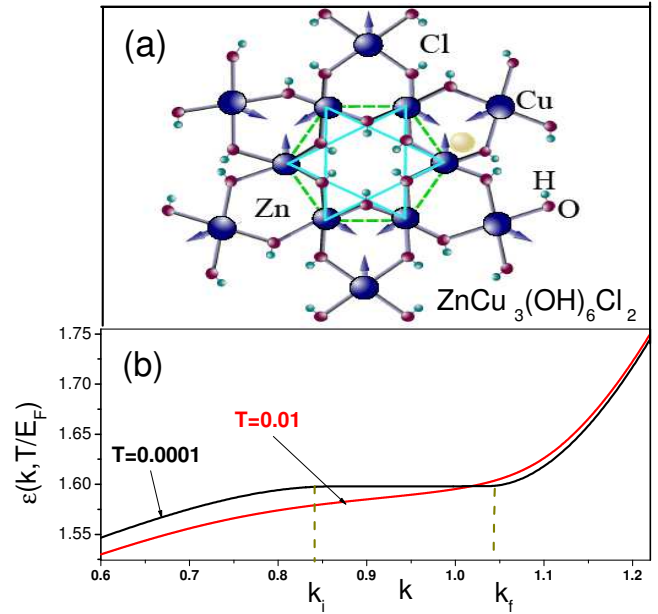}
\end{center}
\vspace*{-0.8cm} \caption{Kagome lattice and flat band. Panel
(a): Crystal structure of $\rm ZnCu_3(OH)_6Cl_2$ along the
hexagonal $c$ axis. The blue spheres are $\rm Cu$ sites. Arrows
show $\rm Cu$ spins in the frustrated configuration. The kagome
planes with frustrated spins, occupying flat band (panel (b)),
are shown by two triangles and highlighted by the dash lines,
displaying the hexagon. Groups $\rm OH$, $\rm Zn$ and $\rm Cl$
atoms are shown schematically. $\rm Zn$ atoms may be situated
below or above $\rm Cu$ planes. Panel (b): Calculated single
particle spectrum $\varepsilon(p/p_F,T/E_F)$ as a function of
dimensionless momentum $k=p/p_F$ and temperature $T/E_F$ with
$E_F$ being Fermi level \cite{prr,book}. The cases of almost
zero ($T=10^{-4}E_F$; red curve) and finite ($T=10^{-2}E_F$;
black curve) temperatures are reported. It is seen that the
dispersionless part of the spectrum vanishes at finite
temperatures making the situation almost indistinguishable from
normal Fermi liquid, while at $T \to 0$  the band becomes almost
flat at $k_i<k<k_f$ where $k_i=p_i/p_F$ and $k_f=p_f/p_F$.}
\label{hrb}
\end{figure}

If the quasiparticles forming QSL are approximately
dispersionless spinons (uncharged fermions), this state of
matter becomes a SCQSL,
\cite{shaginyan:2011,shaginyan:2011:C,book,jltp:2017,jltp:2017o}.
If an insulating chemical compound has 2D lattice, a SCQSL can
emerge as a result of the lattice geometrical frustration. Most
frequently, such frustration results in a dispersionless
topologically protected spectral branch with zero excitation
energy, called flat band. This scenario realizes in
herbertsmithite with its kagome planes
\cite{ks91,green,volovik,vol,vol1,shaginyan:2011} shown in Fig.
\ref{hrb} (a). In this case, SCQSL has its quantum critical
point at the FCQPT. As we have mentioned above, the SCQSL
consists of chargeless  $S=1/2$ spinons having the effective
mass $M^*$. Their momenta, as usual for fermions, reside in a
Fermi sphere with momentum $p_F$. Note, that one frustrated
valence spin is taken away to populate the approximately flat
spinon band, as it is displayed in Fig. \ref{hrb} (b). Such a
behavior strongly resembles that of HF metals whose valence
electrons form conduction flat bands. While the charges $-e$
cannot form any band because of the large charge gap of
herbertsmithite or another geometrically frustrated insulating
magnet. This result is in agreement with the experimental data
\cite{kelly}. Therefore, the above insulating compounds have
physical properties resembling those of HF metals. There is,
however, a notable exception. Namely, while HF metals are good
conductors, the typical insulator does not sustain  the electric
current \cite{shaginyan:2011,shaginyan:2011:C,shaginyan:2012:A}.
Thus, both frustrated insulators and HF metals,having universal
behavior, form a new state of matter \cite{jltp:2017}.

\section{Thermodynamic properties}
If QSL forming quasiparticles are approximately dispersionless
spinons (uncharged fermions), this state of matter becomes a
SCQSL
\cite{shaginyan:2011,shaginyan:2011:C,book,jltp:2017,jltp:2017o}.
As we have mentioned above, the geometrical frustration in
herbertsmithite promotes the SCQSL formation
\cite{shaginyan:2011,shaginyan:2012:A,shaginyan:2011:C}, while
the presence of randomly scattered impurities can facilitate the
frustration, see e. g. \cite{screp-17}.  Note that in real
chemical compounds with many lattice imperfections, the actual
SCQSL emergence point is shifted from the theoretically
predicted FCQPT. This means that the mutual location of the
SCQSL and FCQPT can only be extracted from the experimental
data.

Famous Fermi liquid theory, proposed by Landau (so-called Landau
Fermi liquid (LFL) theory) \cite{land} has been {\it de facto}
the universal tool to describe the itinerant fermionic systems.
It maps the ensemble of strongly interacting electrons in a
solid to the effective quasiparticle gas with not so strong
interaction. In such approach, the excitations are represented
in terms of above quasiparticles so that the low-temperature
properties of the system under consideration depend on the
latter. The quasiparticles possess certain effective mass $M^*$,
which is a parameter of the theory \cite{land,lanl1,PinNoz},
being approximately independent of external stimuli including
temperature, pressure, or an electromagnetic field. The LFL
theory cannot, however, explain the experimental results related
to strong temperature and/or magnetic field dependence of the
effective mass $M^*$, as observed in strongly correlated Fermi
systems \cite{prr,coleman:2002}. At the same time, deviations
from LFL behavior are observed in the vicinity of a FCQPT
\cite{prr,coleman:2002,book}.  Above peculiarities are usually
addressed as non-Fermi-liquid (NFL) properties. They are due to
large effective mass, which becomes actually infinite in the
FCQPT point, see \cite{prr,book} for details.

Let us describe the physical mechanism yielding the temperature
and field dependences of the effective mass of the Landau
quasiparticle $M^*(B,T)$. Again, the key point is that upon
approach to the FCQPT from the LFL regime, the effective mass
becomes strongly external stimuli dependent. Aforementioned
stimuli are the temperature, the magnetic field, the external
pressure $P$ to name a few \cite{prr,book}.  This is indeed a
consequence of an additional (to those of Pomeranchuk,
\cite{lanl1}) instability channel of a normal Fermi liquid. Note
that the new channel is activated when the effective mass
approaches infinity.

To avoid unphysical situations related, for instance, to the
negativity of the effective mass, the system alters the topology
of its Fermi surface \cite{vol1,vol, volovik} so that the
effective mass starts to depend of above external stimuli
\cite{prr,book}. For our studies of the effective mass
properties in SCQSL, we adopt so-called homogeneous Fermi liquid
model. Such description avoids consideration of nonuniversal
(and much irrelevant in our analysis) features like exact
structure of a specific sample \cite{prr,book}. In such model,
the LFL equation for $M^*(B,T)$ reads \cite{land,prr,book}
\begin{eqnarray}
\nonumber \frac{1}{M^*(B,
T)}&=&\frac{1}{M}+\sum_{\sigma_1}\int\frac{{\bf p}_F\cdot{\bf
p}}{p_F^3}F
({\bf p_F},{\bf p}) \\
&\times&\frac{\partial n_{\sigma_1} ({\bf
p},T,B)}{\partial{p}}\frac{d{\bf p}}{(2\pi)^3}. \label{HC1}
\end{eqnarray}
In this expression, $M$ is the free electron mass, $F({\bf
p_F},{\bf p})$ is the Landau interaction function, depending on
$p$ (momentum), $p_F$ (constant Fermi momentum), and
$n_\sigma({\bf p},T,B)$, which is the quasiparticle distribution
function for spin projection $\sigma$. The quasiparticle
interaction $F({\bf p},{\bf p}_1)$, assumed here to be
spin-independent, is phenomenological. Without loss of
generality, here we assume that $F({\bf p},{\bf p}_1)$ is
independent of temperature so that $n_\sigma({\bf p},T)$ has
Fermi-Dirac form
\begin{equation}
n_{\sigma}({\bf p},T)=\left\{ 1+\exp
\left[\frac{(\varepsilon_{\sigma}({\bf
p},T)-\mu_{\sigma})}T\right]\right\} ^{-1},\label{HC2}
\end{equation}
where $\varepsilon_\sigma({\bf p},T)$ is a single-particle
energy spectrum.  Also, $\mu$ is a chemical potential, which is
spin dependent via Zeeman splitting $\mu_{\sigma}=\mu\pm
\mu_BB$, where $\mu_B$ is the Bohr magneton. As usually, the
spectrum $\varepsilon_{\sigma}({\bf p},T)$ is obtained
variationally from the system energy $E[n_{\sigma}({\bf p},T)]$,
\begin{equation} \label{rac}
\varepsilon_{\sigma}({\bf p},T)= \frac{\delta E[n({\bf p})]}{\delta n_{\sigma}}.
\end{equation}

In describing herbertsmithite as a strongly correlated quantum
spin liquid, the choice of the function $F({\bf p},{\bf p}_1)$
is dictated by its possession of FCQPT \cite{prr}.  Thus, the
sole role of the Landau interaction is to drive the system to
the FCQPT point, at which the topology of the Fermi surface is
altered in such a way that the effective mass ceases to be a
constant  parameter obtaining the aforementioned dependence on
external stimuli \cite{prr,book}. Performing the variation
\eqref{rac}, we arrive at the expression for single-particle
energy spectrum
\begin{equation}\label{epta}
\frac{\partial\varepsilon_\sigma({\bf p},T)}{\partial{\bf p}}
=\frac{{\bf p}}{M}-\int \frac{\partial F({\bf p},{\bf
p}_1)}{\partial{\bf p}}n_{\sigma}({\bf
p}_1,T)\frac{d^3p_1}{(2\pi)^3},
\end{equation}

Equations \eqref{HC2} and \eqref{epta} constitute the closed set
to find $\varepsilon_\sigma({\bf p},T)$ and $n_{\sigma}({\bf
p},T)$. In this case the effective mass enters through the
expression $p_F/M^*=\partial\varepsilon(p)/\partial p|_{p=p_F}$.
At the FCQPT point, the analytical solution of Eq.~\eqref{HC1}
is possible \cite{prr,book}. Namely, in contrast to LFL approach
with $M^*$ being a constant parameter, here at zero magnetic
field, $M^*$ becomes temperature dependent. Latter feature
comprises the strong deviation from LFL picture, determining the
NFL regime
\begin{equation}
M^*(T)\simeq a_TT^{-2/3}.\label{MTT}
\end{equation}
At elevated temperatures, the system undergoes a transition to
the LFL region of the phase diagram, and being subjected to the
magnetic field, exhibits the behavior
\begin{equation}
M^*(B)\simeq a_BB^{-2/3}\label{MBB}
\end{equation}
of the effective mass.

The introduction of "internal" (or natural) scales greatly
simplifies analysis of the problem under consideration.  We
first observe that near the FCQPT, the effective mass $M^*(B,T)$
(i.e. the solution of Eq.~\eqref{HC1}) has a maximum $M^*_M$ at
a temperature $T_{M}\propto B$ \cite{prr,book}. This means, that
it is convenient to to measure the effective mass and
temperature in the units $M^*_M$ and $T_{M}$ respectively. Thus
we arrive at normalized effective mass $M^*_N=M^*/M^*_M$ and
temperature $T_N=T/T_{M}$.  Near FCQPT, the dependence
$M^*_N(T_N)$ can be rendered as a universal interpolating
function \cite{prr,book}. This function describes the transition
from LFL to NFL states, given by Eqs. \eqref{MBB} and
\eqref{MTT} and presents the universal scaling behavior of
$M^*_N$ \cite{prr,book}
\begin{equation}M^*_N(y)\approx c_0\frac{1+c_1y^2}{1+c_2y^{8/3}}.
\label{UN2}
\end{equation}
Here $y=T_N=T/T_{M}$ and $c_0=(1+c_2)/(1+c_1)$, where $c_1$ and
$c_2$ are free parameters. The magnetic field $B$ enters
Eq.~\eqref{HC1} only in the combination $\mu_BB/T$, making
$T_{M}\sim \mu_BB$. It follows from Eq.~\eqref{UN2} that
\begin{equation}
\label{TMB2} T_M\simeq a_1\mu_BB,
\end{equation}
where $a_1$ is a dimensionless quantity. In this case,  the
variable $y$ becomes $y=T/T_{M}\sim T/\mu_BB$. The equation
\eqref{TMB2} permits to assert Eq.~\eqref{UN2} gives the scaling
properties of temperature and magnetic field dependence of the
effective mass. This implies, in turn, that at different fields
$B$,  the curves $M^*_{N}$  form a single one as a function of
the normalized variable $y=T/T_M$. As $T$ and $B$ enter
symmetrically in Eq.~\eqref{UN2}, it also manifests the scaling
behavior of $M^*_{N}(B,T)$ as a function of $T/B$:
\begin{equation}
\label{TMB} T_N=\frac{T}{T_M}= \frac{T}{a_1\mu_BB}\propto
\frac{T}{B}\propto \frac{B}{T}.
\end{equation}

\begin{figure}[!ht]
\begin{center}
\includegraphics [width=0.47\textwidth]{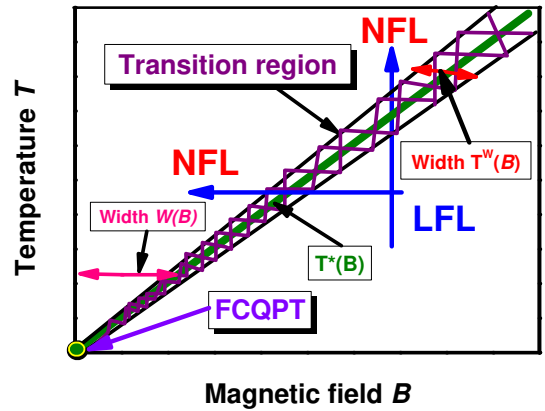}
\end{center}
\caption{(Color online). Schematic SCQSL phase diagram in the
"temperature-magnetic field" representation.   Magnetic field
$B$ is the independent variable (control parameter). Vertical
and horizontal arrows show LFL-NFL and NFL-LFL transitions at
fixed $B$ and $T$, respectively. The dependences $M^*(T)$ and
$M^*(B)$ are given by Eqs.~\eqref{MTT} and \eqref{MBB},
respectively. The hatched area represents the transition region
at $T=T^*(B)$, see Eq.~\eqref{BMT}. The transitions occur
according to the directions of the blue arrows. The solid line
in the hatched area represents the function $T^*(B)\simeq
T_M(B)$ given by Eq.~\eqref{TMB2}. The functions $W(B)\propto
T\propto T^*$ and $T^W(B)\propto T\propto T^*$ shown by
two-headed arrows define the width of the NFL state and the
transition area, respectively. At FCQPT indicated by the arrow,
the effective mass $M^*$ diverges and both $W(B)$ and $T^W(B)$
tend to zero.}\label{fig0}
\end{figure}

The schematic phase diagram is portrayed  in Fig.~\ref{fig0}. We
assume for simplicity that at $T=0$ and $B=0$ the system is near
FCQPT. The external magnetic field and temperature are indeed
the parameters, controlling the system position on the phase
diagram relatively to FCQPT point. The same parameters are
responsible for the transitions between the NFL and LFL regions
of the phase diagram, see the arrows in Fig.~\ref{fig0}.
Horizontal arrow corresponds to fixed temperatures so that the
system motion along this arrow from NFL to LFL region
corresponds to magnetic field increase. On the contrary, the
vertical arrow fixes magnetic field so that the motion along it,
signifies the elevated temperatures. The shaded area reports the
region, where the NFL state transits to weakly polarized LFL
one. The temperature $T^*(B)$ of the transition is defined by
the expression
\begin{equation}
\label{BMT}
T^*(B)\simeq T_M(B),
\end{equation}
which directly follows from Eq.~\eqref{TMB2}. The line
Eq.~\eqref{BMT} is indeed the function $T^*\propto \mu_BB$, so
that width $W(B)$ of the NFL region is proportional to the
temperature. It can be shown similarly that the transition
region width $T^W(B)$ is also $\propto T$  \cite{prr,book}. We
note here that in essence the transition region represents the
crossover between LFL and NFL phases. In our case, the NFL phase
is formed by quasiparticles that occupy the so-called fermion
condensate (FC) state, in analogy to the Bose condensate for
particles obeying Bose-Einstein statistics.  (See
Ref.~\cite{prr,book} for a comprehensive explanation.)  In a
"pure" FC state, all fermions (quasiparticles) having momenta in
a finite interval embracing the Fermi surface have energies
pinned to the chemical potential. In reality this state cannot
be reached because of the Nernst theorem \cite{annals}, and the
NFL features arise from "traces'' of the FC state manifested at
finite temperatures. Also, at low but finite temperatures, the
magnetic field acts to suppress NFL behavior (i.e., the "FC
traces") and, on growing sufficiently strong, restores the LFL
phase.  On the other hand, thermal fluctuations destroy LFL
behavior and generate NFL features related to the FC state.
\begin{figure} [! ht]
\begin{center}
\includegraphics [width=0.47\textwidth]{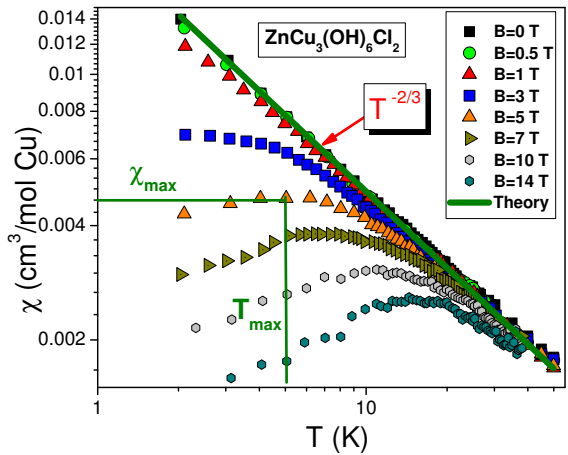}
\end{center}
\vspace*{-0.8cm} \caption{(Color online) The dependence
$\chi(T)$, measured experimentally on $\rm ZnCu_3(OH)_6Cl_2$,
taken  from Ref.~\cite{herb3}. Magnetic fields are coded by
symbols and reported in the legend.  The example of $\chi_{\rm
max}$ and $T_{\rm max}$ at $B=3$ T is shown. The result of
calculation at $B=0$ (full line) shows the dependence
$\chi(T)\propto T^{-2/3}$
\cite{shaginyan:2011,shaginyan:2013:D,book}.} \label{fig01}
\end{figure}
In fact, at $T=0$ the FC state is represented by the
superconducting state with the superconducting gap $\Delta=0$,
while the superconducting order parameter
$\kappa=\sqrt{n(p)(1-n(p))}$ is finite in the region $(p_i-p_f)$
\cite{khod1994,prr,pccp}, for in the region $n(p)<1$.

\begin{figure} [! ht]
\begin{center}
\includegraphics [width=0.47\textwidth]{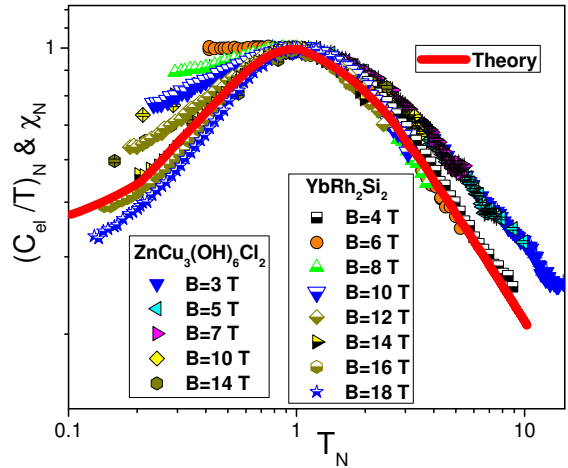}
\end{center}
\caption{Normalized susceptibility $\chi_N=\chi/\chi_{\rm
max}=M^*_N$ versus normalized temperature $T_N$ (see
Eq.~\eqref{UN2}) of $\rm ZnCu_3(OH)_6Cl_2$ \cite{herb3} as
collected in Fig.~\ref{fig01}. We normalize specific heat
$(C_{el}/T)_N=M^*_N$ taken from the experimental results on
$C_{el}/T$ in $\rm YbRh_2Si_2$ in magnetic fields
$B$ shown in Fig. \ref{fig45} (b) \cite{steg1}. The
corresponding values of $B$ (legends) are coded by symbols.
The full curve marks our theoretical calculations
at $B\simeq B^*$ when the quasiparticle band
is fully polarized. \cite{book,shaginyan:2011:C}.}\label{fig4_1}
\end{figure}

\begin{figure} [! ht]
\begin{center}
\includegraphics [width=0.47\textwidth]{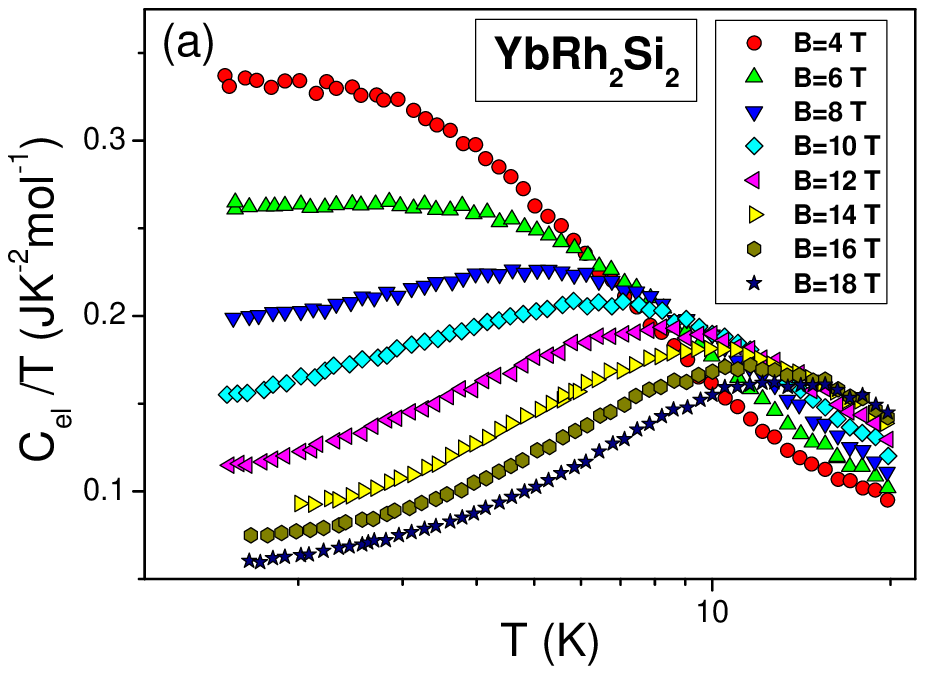}
\includegraphics [width=0.47\textwidth]{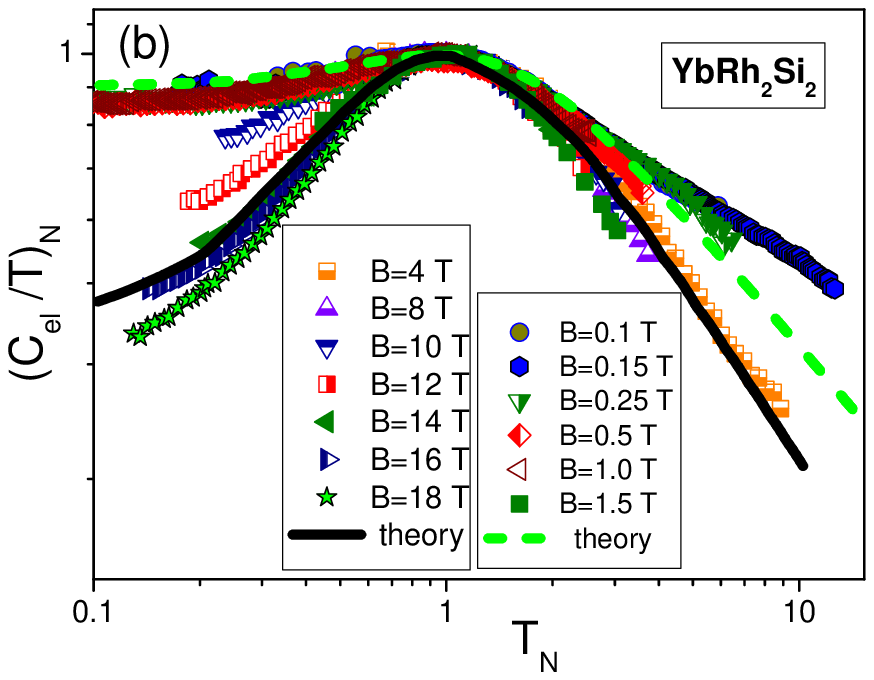}
\end{center}
\caption{(Color online) Electronic specific heat $C_{el}/T$ of
archetypical HF metal $\rm YbRh_2Si_2$. Panel (a) reports
the temperature dependence of the electronic specific heat
$C_{el}/T$ of $\rm YbRh_2Si_2$ at high magnetic fields
\cite{steg1} shown in the legend. The electronic specific heat
$C_{el}/T$ of $\rm YbRh_2Si_2$ strongly resembles $C_{\rm
mag}/T$ of $\rm ZnCu_3(OH)_6Cl_2$ shown in Fig. \ref{fig02}.
Panel (b): The normalized specific heat $C_{el}/T$ at high
\cite{steg1} and low magnetic fields \cite{oesb} extracted from
the specific heat ($C/T$) measurements on the $\rm YbRh_2Si_2$.
The low-field calculations are depicted by the short dash curve
tracing the scaling behavior of $M^*_N$, see Eq. \eqref{UN2}.
Our high-field calculations (solid line) are taken at high
magnetic field $B$ at which the quasiparticle band becomes fully
polarized \cite{shag2011}.}\label{fig45}
\end{figure}

\begin{figure} [! ht]
\begin{center}
\includegraphics [width=0.47\textwidth]{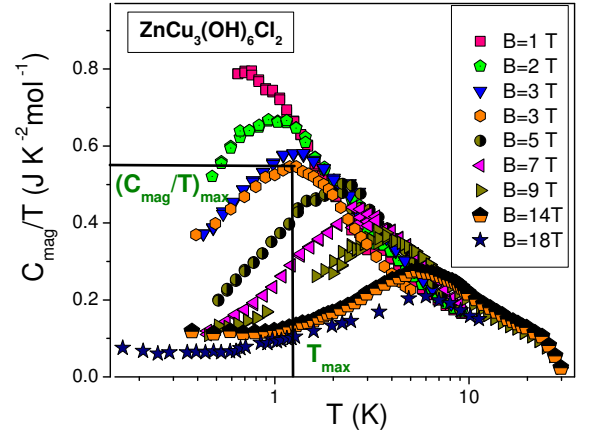}
\end{center}
\vspace*{-0.8cm} \caption{(Color online) Specific heat $C_{\rm
mag}/T$ measured on powder \cite{helt,herb2} and single-crystal
\cite{herb,t_han:2012,t_han:2014} herbertsmithite  samples as
function of temperature. Magnetic fields (legend) are coded by
symbols. It is clearly seen that powder and single-crystal
samples has approximately the same $C_{\rm mag}/T$. The example
of $C_{\rm mag}/T_{\rm max}$ and $T_{\rm max}$ at $B\simeq 3$ T
is shown.} \label{fig02}
\end{figure}

\begin{figure} [! ht]
\begin{center}
\includegraphics [width=0.47\textwidth]{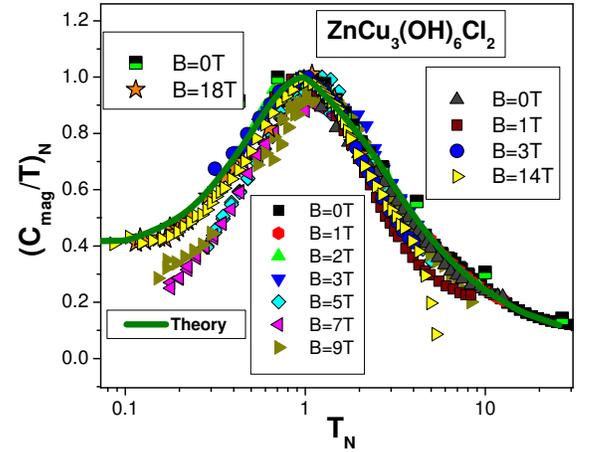}
\end{center}
\vspace*{-0.8cm} \caption{(Color online) Normalized specific
heat $(C_{\rm mag}/T)_N$ versus normalized temperature $T_N$ at
at different magnetic fields (legend) \cite{hfliq,book}
extracted from specific heat $C_{\rm mag}/T$ measured on powder
\cite{helt,herb2} and single-crystal
\cite{herb,t_han:2012,t_han:2014} herbertsmithite samples, see
Fig. \ref{fig02}. The theoretical result from
Refs.~\cite{shaginyan:2011,book}, represented by the solid
curve, traces the scaling behavior of the normalized specific
heat.} \label{fig03}
\end{figure}

To examine the impurity model and possible gap in the spinon
single particle spectrum we consider the measured properties of
herbertsmithite magnetic susceptibility $\chi$. It is seen from
Fig.~\ref{fig01} that the magnetic susceptibility diverges as
$\chi(T)\propto T^{-2/3}$ at $B\leq 1$ T (full line).  For
weakly interacting impurities it has been suggested that at low
temperatures, the dependence $\chi(T)$ can be approximated by a
Curie-Weiss law \cite{Han,Han11,sc_han}, i.e., $\chi_{\rm
CW}(T)\propto 1/(T+\theta)$, where $\theta$ is a Curie
temperature, which turns out to be very small. At the same time,
with respect to firmly established behavior $\chi(T)\propto
T^{-2/3}$ at low $B$, the above Curie-Weiss approximation is in
discord with both theory
\cite{shaginyan:2011,shaginyan:2011:C,book} and experiment
\cite{herb3}. Moreover, as seen in Fig.~\ref{fig4_1}, the
normalized spin susceptibility $\chi_N$ behaves like the
normalized specific heat $C_{el}/T$ extracted from measurements
on $\rm YbRh_2Si_2$ in high magnetic fields \cite{steg1} and
displayed in Fig. \ref{fig45} (a). Note that $C_{el}/T$
displayed in Fig. \ref{fig45} (a) coincides  approximately with
$C_{mag}(B,T)/T\simeq C_{el}(B,T)/T$ shown in Fig. \ref{fig02}
pointing to similarity of the electron and spinon Fermi spheres.
Fig. \ref{fig45} (b) reports the normalized specific heat
$(C_{el}/T)_N$ extracted from the data \cite{steg1,oesb}. It is
seen that at low magnetic fields the electronic system is not
polarized as it does at high ones (black line)
\cite{shag2011,book}. We note that the same behavior is
exhibited by both $\chi_N$ shown in Fig. \ref{fig02} and
$(C_{mag}/T)_N$ shown in Fig. \ref{fig03}. These facts support
our conclusion that SCQSL of $\rm ZnCu_3(OH)_6Cl_2$ behaves like
both the HF electron liquid of $\rm YbRh_2Si_2$ and the
insulator $\rm {Cu(C_4H_4N_2)(NO_3)_2}$
\cite{sc2016,ann_ph2016}.

The above observations point to a possible absence of the spin
gap in $\rm ZnCu_3(OH)_6Cl_2$, for it is impossible to
definitely separate the contributions coming from the impurities
from that of kagome planes. The impurity model implies that the
intrinsic spin susceptibility of the kagome plane is decomposed
as $\chi_{\rm kag}(T)=\chi(T) -\chi_{\rm CW}(T)$, leading to
$\chi_{\rm kag}(T\to0)\to 0$ and the result about existence of a
putative gap \cite{Han11}. This shows that there is a problem
with the impurity model as it cannot describe the firmly
established behavior $\chi(T)\propto T^{-2/3}$ \cite{herb3}.
Thus, to explain the observed behavior of $\chi$, one should
view the impurity ensemble embedded in the kagome host crystal
lattice as an {\it integral} system
\cite{prr,shaginyan:2011,shaginyan:2012:A,shaginyan:2011:C,
shaginyan:2013:D,comm,hfliq,book,sc2016,ann_ph2016} that acts
coherently at nanoscale to produce additional frustration of the
kagome planes, so as to make QSL robust at lower temperatures.
Based on this analysis, we predict that QSL of quantum magnets
can be stabilized introducing a random distribution of
impurities. We also note that artificially or spontaneous
introducing impurities into $\rm ZnCu_{3}(OH)_6Cl_2$ can both
stabilize its QSL and simplify its chemical preparation.

\begin{figure} [! ht]
\begin{center}
\includegraphics [width=0.47\textwidth]{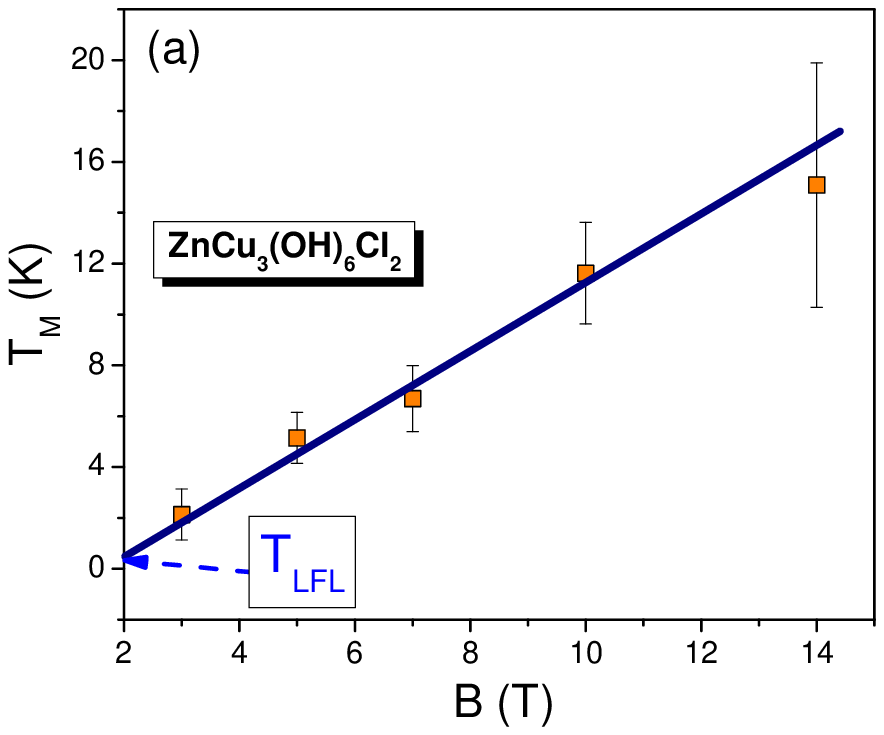}
\includegraphics [width=0.47\textwidth]{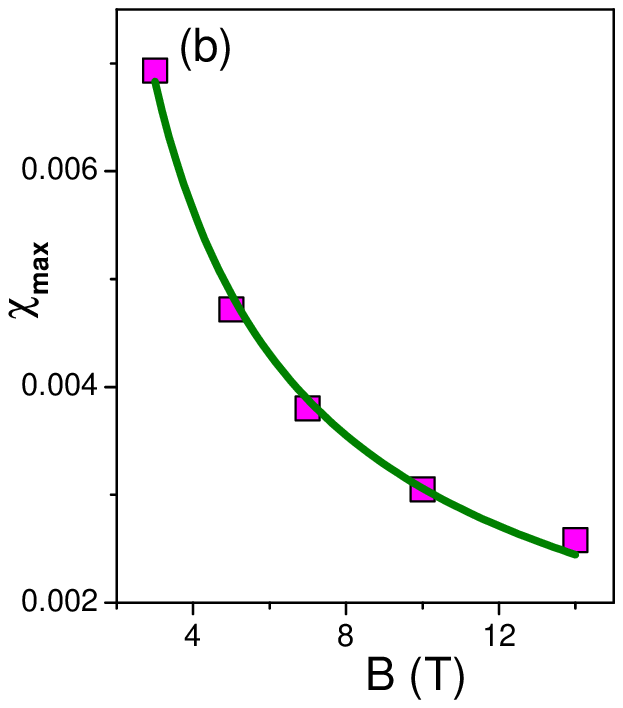}
\end{center}
\vspace*{-0.8cm} \caption{(Color online) Panel (a): The
temperatures $T_{M}(B)$ at which the maxima of
$\chi_{max}(B)\propto M^*_M(B)$ (see Fig. \ref{fig01}) occur.
The solid line represents the function $T_{M}\propto B$, see Eq.
\eqref{TMB2}. Panel (b): the maxima $\chi_{max}(B)$ of $\chi(T)$
versus $B$ (see Fig. \ref{fig01}). The solid curve is given by
Eq. \eqref{MBB}.} \label{fig313}
\end{figure}

It is instructive to plot the functions $T_M(B)$ and
$\chi_M(B)\propto M^*_M(B)$ of $\rm ZnCu_3(OH)_6Cl_2$. These
functions are displayed in Fig. \ref{fig313} with the data
extracted from experimental facts \cite{helt}. It is seen from
Fig. \ref{fig313} (a) that behavior of $T_M(B)\propto B$ is in
accordance with Eq. \eqref{MBB}. It is also seen from Fig.
\ref{fig313} (a) that $\rm ZnCu_3(OH)_6Cl_2$ is not located
exactly at FCQPT, for $T_M(B=0)\simeq 0.4$ K and the system
demonstrate the LFL behavior at $T\leq 0.4$ K \cite{helt}.
Similarly, it is seen from Fig. \ref{fig313} (b) that
$\chi_{max}(B)\propto M^*_M(B)$ in accordance with Eqs.
\eqref{MBB} and \eqref{UN2}. Thus, the effect of the impurities
on QSL can be analyzed basing on Fig. \ref{fig313}: the LFL
temperature $T_{LFL}$ is growing if QSL is shifted by impurities
from FCQPT, while the gap is absent. It also follows from Fig.
\ref{fig313} that $\rm ZnCu_3(OH)_6Cl_2$ exhibit typical
behavior of HF metals under the application of magnetic field
\cite{prr,book}. As a result, we conclude that impurities form
the single integral entity with kagome lattice at nanoscale.

The impurity model has been utilized by the authors of
Ref.~\cite{Han} to derive an intrinsic scattering measure
$S_{\rm kag}(\omega)=S_{\rm tot}(\omega) -aS_{\rm imp}(\omega)$.
In the latter expression, $S_{\rm tot}(\omega)$ is a total (i.e.
host lattice plus impurities) scattering rate, while $S_{\rm
imp}(\omega)$ is the impurity contribution with fitting
parameter $a$. On finding that $S_{\rm kag}(\omega)$ goes to
zero as $\omega < 0.7$ meV (see Fig.~4(b) of Ref.~\cite{Han})
they assert the existence of a gap. However, as we have
demonstrated above, such a procedure gives just a spurious gap.
Indeed, latter conclusion relies completely on the assumption
about weak inter-impurity interaction, which is unjustified
empirically since the subtraction hypothesis is negated by the
experimental behavior summarized in Fig.~\ref{fig01}

We now examine the impurity model in further details and put its
conclusion about spin gap existence under scrutiny. It is seen
from Fig.~\ref{fig01} that normal low-temperature Fermi liquid
properties of the magnetic susceptibility $\chi$ is confirmed
experimentally at least for $B\geq 3$ T.  In such strong
magnetic field and low temperatures the impurity spins are
aligned completely along the field direction.  They do not have
degrees of freedom, permitting them to demonstrate Curie-Weiss
behavior. Thus, assuming that impurity spins are fully aligned
along magnetic field direction and hence do not contribute to
$\chi$, one remains with $\chi_{\rm kag}(T) =\chi(T)$. Similar
properties of the heat capacity follow from Fig.~\ref{fig02}. In
the above magnetic fields, $C_{\rm mag}/T$ also demonstrates
ordinary Fermi liquid behavior. This shows that at low
temperatures and $B\geq 3$ T, the impurities give minute
contributions to $\chi$ and $C_{\rm mag}/T$. It becomes obvious
that the main contribution in such case comes from the host
kagome lattice \cite{Han,Han11,sc_han}. Moreover, impurity model
implies that both $\chi(T)$ and $C_{\rm mag}(T)/T$ approach zero
at $T \to 0$ and $B\geq 3$ T. It is clear from
Figs.~\ref{fig01}, \ref{fig4_1}, \ref{fig02}, and \ref{fig03},
that this is not true. Namely, both  $\chi$ and $C_{\rm mag}/T$
do not go to zero at  $T\to0$ up to $B\sim 14$ T. Moreover, the
scaling properties of $C_{\rm mag}/T$ from Fig.~\ref{fig03} and
the behavior of both $T_M(B)$ and $\chi_{max}(B)$ displayed in
Fig. \ref{fig313} confirm clearly the lack of a gap. The results
of $C_{\rm mag}$ measurements
\cite{helt,herb2,herb,t_han:2012,t_han:2014} are the same for
powder and bulk samples, see Fig.~\ref{fig02}.

The aforementioned experimental results are consistent with the
hypothesis that the vast majority of physical properties of
herbertsmithite are due to stable SCQSL. First, there is no
substantial gap in the spinon excitation spectrum. Moreover, the
application of high magnetic fields 18 T does not trigger such
gap. This implies that the impurity model is inadequate.  These
findings are in conformity with recent results of the
measurements, stating that the low-temperature plateau in the
local susceptibility identifies the spin-liquid ground state as
a gapless one \cite{zorko}, while recent theoretical analysis
confirms the absence of a gap \cite{Normand}.

\section{Dynamic, relaxation and heat transport properties}

\begin{figure} [! ht]
\begin{center}
\includegraphics [width=0.47\textwidth]{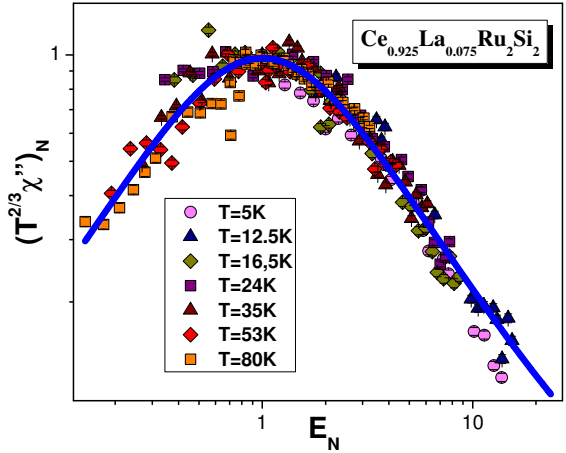}
\end{center}
\caption{(Color online) Normalized dynamic spin susceptibility
$(T^{2/3}\chi'')_N$, showing scaling properties. The data are
taken from those on the HF metal $\rm
Ce_{0.925}La_{0.075}Ru_2Si_2$ \cite{knafo:2004}. Full line is
defined by Eq.~\eqref{SCHIN}
\cite{shaginyan:2012:A}.}\label{fig04}
\end{figure}

\begin{figure} [! ht]
\begin{center}
\includegraphics [width=0.47\textwidth]{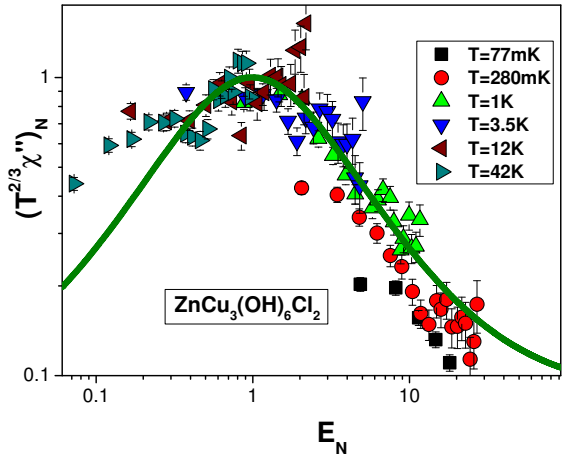}
\end{center}
\caption{(Color online)  Same as in Fig. \ref{fig04} but for
herbertsmithite $\rm ZnCu_3(OH)_6Cl_2$ \cite{herb3}. Full line
is once more defined by Eq.~\eqref{SCHIN}
\cite{shaginyan:2012:A,book}.}\label{fig05}
\end{figure}

\begin{figure} [! ht]
\begin{center}
\includegraphics [width=0.47\textwidth]{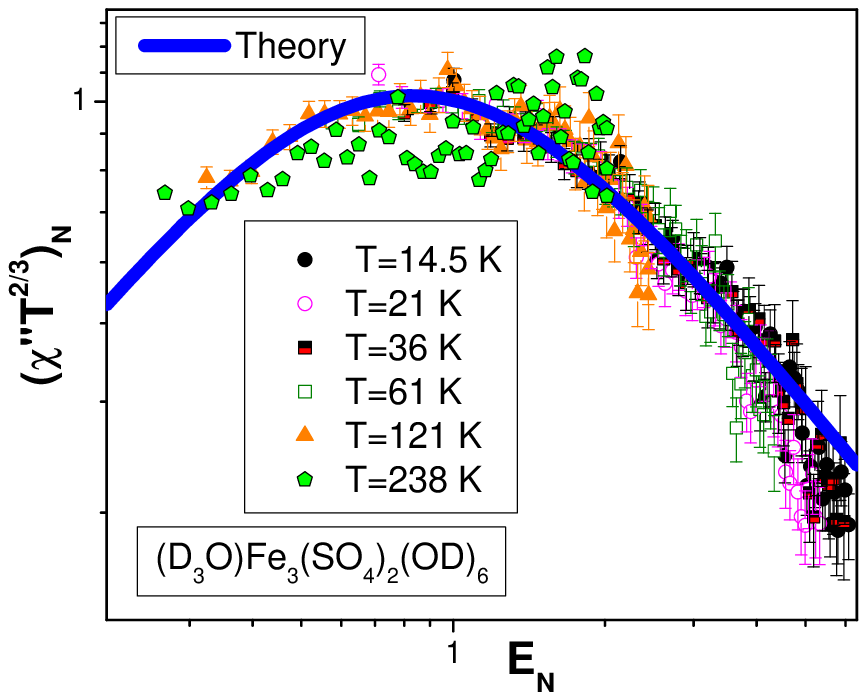}
\end{center}
\caption{(Color online) Same as in Fig. \ref{fig04} but for
deuteronium jarosite $\rm (D_3O)Fe_3(SO_4)_2(OD)_6$
\cite{faak:2008}.  Full line is once more defined by
Eq.~\eqref{SCHIN} \cite{shaginyan:2012:A,book}.}\label{fig06}
\end{figure}

\begin{figure} [! ht]
\begin{center}
\includegraphics [width=0.47\textwidth]{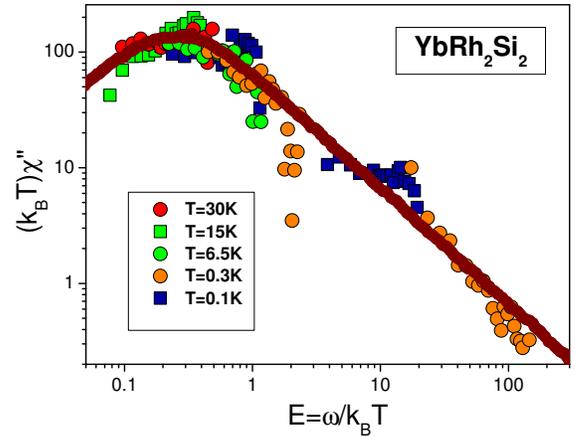}
\end{center}
\caption{(Color online) Normalized dynamic spin susceptibility
$T\chi''$ against $E=\omega/k_BT$. Data are taken from those on
$\rm YbRh_2Si_2$ \cite{stock}. Full line is defined by
Eq.~\eqref{SCHIT} \cite{shaginyan:2012:A,book}.}\label{fig09}
\end{figure}

Neutron scattering measurements is one more confirmation of our
hypothesis validity. Recently, the dynamic spin susceptibility
$\chi({\bf q},\omega,T) =\chi{'}({\bf
q},\omega,T)+i\chi{''}({\bf q},\omega,T)$ has been measured by
neutrons scattering as a function of $\bf q$ (momentum),
$\omega$ (frequency) and $T$. At low temperatures, the results
are consistent with the idea that these quasiparticles are
spinons, which form an approximately flat band \cite{Han:2012}.

The imaginary part $\chi''(T,\omega_1)$ satisfies the equation
\cite{shaginyan:2012:A,book}
\begin{equation}\label{SCHII}
T^{2/3}\chi''(T,\omega_1)\simeq\frac{a_1\omega_1}{1+a_2\omega_1^2},
\end{equation}
where $a_1$ and $a_2$ are constants and
$\omega_1=\omega/(T)^{2/3}$. Equation ~\eqref{SCHII}
demonstrates that $T^{2/3}\chi''(T,\omega_1)$ has a maximum
$(T^{2/3}\chi''(T,\omega_1))_{\rm max}$ at some $\omega_{\rm
max}$. Equation \eqref{SCHII} describes the scaling properties
of $\chi'' T^{0.66}$ established in measurements in
Ref.~\cite{herb3}. As in Eq.~\eqref{UN2}, we introduce the
dimensionless function
$(T^{2/3}\chi'')_{N}=T^{2/3}\chi''/(T^{2/3}\chi'')_{\rm max}$
and the (dimensionless) variable $E_N=\omega_1/\omega_{\rm
max}$. In this case, Eq.~\eqref{SCHII} is modified to read
\begin{equation}\label{SCHIN}
(T^{2/3}\chi'')_N\simeq\frac{b_1E_N}{1+b_2E_N^2},
\end{equation}
where $b_1$ and $b_2$ are fitting parameters.They should be
chosen from the condition that the right hand side of
Eq.~\eqref{SCHIN} should reach the maximum at $E_N=1$. This
means that the expression $(T^{2/3}\chi'')_{N}=T^{2/3}\chi''
/(T^{2/3}\chi'')_{\rm max}$ exhibits scaling as a function of
$E_N $ \cite{shaginyan:2012:A,book}. In other words, we find
that $B^{2/3}\chi''(\omega)$ (cf.\ Eq.~\eqref{MBB}) exhibits the
scaling behavior with $E_N=\omega_1/\omega_{\rm max}$:
\begin{equation}\label{SCHB}
(B^{2/3}\chi'')_N\simeq\frac{d_1E_N}{1+d_2E_N^2},
\end{equation}
Similarly, $d_1$ and $d_2$ are fitting parameters chosen from
the condition that the function $(B^{2/3}\chi'')_{N}=1$ at
$E_N=1$. In the FCQPT point, the discussed scaling is valid
almost to $T=0$.

Figure~\ref{fig04} reports the function $(T^{2/3}\chi'')_{N}$
extracted from neutron-scattering measurements on the HF metal
$\rm Ce_{0.925}La_{0.075}Ru_2Si_2$ \cite{knafo:2004}. The
corresponding data collected on two systems, $\rm
ZnCu_3(OH)_6Cl_2$ \cite{herb3} and $\rm
(D_3O)Fe_3(SO_4)_2(OD)_6$ \cite{faak:2008}, are reported in
Figs.~\ref{fig05} and \ref{fig06}. The Figures show pretty good
agreement between theoretical \cite{shaginyan:2012:A} (solid
curves) and experimental results on all considered chemical
compounds over almost three decades in the scaled variable
$E_N$.  Hence $(T^{2/3}\chi'')_{N}$ also has scaling properties.
This shows that the spinons in both $\rm ZnCu_3(OH)_6Cl_2$ and
$\rm (D_3O)Fe_3(SO_4)_2(OD)_6$ demonstrate the same itinerant
behavior as the conduction electrons in the heavy-fermion
compound $\rm Ce_{0.925}La_{0.075}Ru_2Si_2$. The detection of
above itinerant behavior is very important as it shows
convincingly the presence of a gapless SCQSL in herbertsmithite
\cite{shaginyan:2012:A,shaginyan:2011:C,book}.

Under the assumption that a fermion condensate (FC) is present
in the above HF metals, the imaginary part $\chi''(T,\omega)$ of
the susceptibility reads \cite{book,prr}
\begin{equation}\label{SCHIT}
T\chi''(T,\omega)\simeq\frac{a_5E}{1+a_6E^2},
\end{equation}
where $E=\omega/k_BT$ while $a_5$ and $a_6$ are adjustable
parameters. Expression \eqref{SCHIT} shows that
$T\chi''(T,\omega)$ is a function of a single variable
$E=\omega/k_BT$. In this case,  the expressions \eqref{SCHIN}
and \eqref{SCHIT} define two two types of scaling in
$\chi''(\omega,T)$. The dynamic susceptibility $(T\chi'')$ taken
from inelastic neutron scattering measurements on the HF metal
$\rm YbRh_2Si_2$ \cite{stock} is reported in Fig.~\ref{fig09}.
The scaling in $(T\chi'')$ is clearly seen in both this function
and variable $E$. This confirms both the feasibility of
Eq.~\eqref{SCHIT} and the similarity of the HF metals and
frustrated magnets properties. The scaled data acquired in
measurements on such structurally different strongly correlated
systems as $\rm ZnCu_3(OH)_6Cl_2$, $\rm
Ce_{0.925}La_{0.075}Ru_2Si_2$, $\rm (D_3O)Fe_3(SO_4)_2(OD)_6$,
and $\rm YbRh_2Si_2$ merge in a single curve over almost three
decades in the scaled variables, thus confirming that these
strongly correlated Fermi-systems exhibit universal scaling
behavior and symptomatic of the existence of a new state of
matter \cite{book,front}. This observation strengthens the
credibility of FC approach as it can reliably explain the
experimental data concerned and has demonstrable predictive
power \cite{book,front,jltp:2017}.

Indeed, it is apparent from Figs.~\ref{fig04}, \ref{fig05},
\ref{fig06}, and \ref{fig09} that the calculations within FC
approach are in conformity with th experimental results,
providing strong evidence that SCQSL is the underlying mechanism
defining the properties of $\rm ZnCu_3(OH)_6Cl_2$ and $\rm
(D_3O)Fe_3(SO_4)_2(OD)_6$. We conclude that the spin gap
apprehension is rather artificial construction, which may
contradict the acquired experimental knowledge about the $\rm
ZnCu_{3}(OH)_6Cl_2$ properties. The consistency of this
description based on a FCQPT as the driving mechanism for the
properties of herbertsmithite at low temperature can also be
taken as strong evidence against the spin gap existence in this
system.  Note, that the presence or absence of a spin gap is not
that important as it does not define the physical properties of
$\rm ZnCu_3(OH)_6Cl_2$.  In particular, neither $\chi$ nor
$C_{\rm mag}/T$ vanishes in the high magnetic fields,
eliminating the contribution coming from the impurities (see
Figs.~\ref{fig01}, \ref{fig4_1}, \ref{fig02}, and \ref{fig03}).
Moreover, Fig.~ \ref{fig05} shows that such a gap does not
contribute to the imaginary part of the magnetic susceptibility,
which does not vanish at the lowest temperatures. Rather, all
above physical properties are due to the SCQSL. This fact can be
tested by magnetic field heat transport measurements similar to
organic insulators $\rm EtMe_3Sb[Pd(dmit)_2]_2$ and $\rm
\kappa-(BEDT-TTF)_2Cu_2(CN)_3$
\cite{yamashita:2010,yamashita:2012,shaginyan:2013:D}. Heat
transport measurements are notably important as they probe the
SCQSL excitations in $\rm ZnCu_3(OH)_6Cl_2$. They can also
detect the itinerant spinons which are the primary reason for
heat transport. It is obvious, that the above heat transport
cannot do without phonons. On the other hand, the phonon
contribution is barely affected by the magnetic field. To
summarize, we anticipate that magnetic field dependence of heat
transport measurements may  be an important step forward to
identify the SCQSL nature in $\rm ZnCu_3(OH)_6Cl_2$
\cite{shaginyan:2011:C,shaginyan:2013:D,book}.

Note that of we set the electronic charge to zero, the SCQSL in
herbertsmithite becomes identically similar to the itinerant
electrons ensemble in HF metals. In this case, the SCQSL thermal
resistivity $w$ reads
\cite{shaginyan:2011:C,shaginyan:2013:D,book}
\begin{equation}\label{wr}
w-w_0=W_rT^2\propto\rho-\rho_0\propto(M^*)^2T^2,
\end{equation}
where $W_r{T^{2}}$comes from spinon-spinon scattering, which is
similar to the contribution $AT^2$ from electron-electron
scattering to charge transport. Here $M^*$ is the effective mass
and $\rho$ is the longitudinal magnetoresistivity (LMR), while
$w_0$ and $\rho_0$ denote the residual thermal and electric
resistivity, respectively.
\begin{figure} [! ht]
\begin{center}
\includegraphics [width=0.52\textwidth]{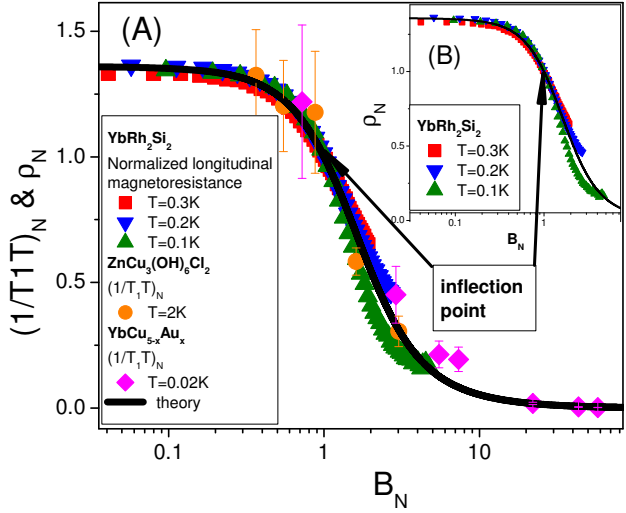}
\end{center}
\caption{(Color online) Main panel (A). Magnetic field
dependence of normalized spin-lattice relaxation rate
$(1/T_1T)_N$. Data for different substances are shown by
different markers: solid squares correspond to  $\rm
ZnCu_3(OH)_6Cl_2$ \cite{imai} and solid triangles - to $\rm
YbCu_{5-x}Au_{x}$ at $x=0.4$ \cite{carr}. The normalization is
done in the inflection point, shown by the arrow. Inset (B).
Same as in main panel (a), but for the normalized
magnetoresistance $\rho_N$, for $\rm YbRh_2Si_2$ at different
temperatures \cite{gegmr} (legend) coded by symbols. The result
of calculations is shown in both panels by the full line,
defining the scaling behavior of $W_r\propto(M^*)^2$ (see
Eqs.~\eqref{wr} and \eqref{WT}).}\label{T12}
\end{figure}
Finally, we consider the magnetic field influence on the
spin-lattice relaxation rate $1/(T_1T)$. Fig.~\ref{T12} A shows
the normalized quantity $1/(T_1T)_N$ as a function of magnetic
field. It shows that at magnetic field increase, $1/(T_1T)$
decays progressively. Contrary to the previous cases, where the
normalization has been fulfilled in the point of maximum, here
we normalize our curves in the inflection point, $B=B_{\rm
inf}$, marked by the arrows in main panel and inset. The same
procedure has been performed with the normalized
magnetoresistance shown in the inset to Fig.~\ref{T12}. The
relation $1/(T_1T)_N\propto(M^*)^2$ shows that our system
located near its QCP would exhibit similar behavior of
$1/(T_1T)_N$ \cite{prr,shaginyan:2011:C,shaginyan:2013:D,book}.
Significantly, Fig.~\ref{T12} A shows that the normalized
spin-lattice relaxation rate of herbertsmithite \cite{imai} and
HF metal $\rm YbCu_{5-x}Au_{x}$ \cite{carr} do exhibit the same
behavior.  As it is seen from Fig.~\ref{T12} A for $B\leq B_{\rm
inf}$ (or $B_N\leq1$) the quantity $1/(T_1T)_N$ is almost
magnetic field independent. At the same time, at elevated
magnetic fields it decays
\cite{prr,shaginyan:2011:C,shaginyan:2013:D,book} according to
\begin{equation}\label{WT}
W_r\propto{ 1/(T_1T)_N}=\rho_N\propto(M^*)^2\propto B^{-4/3}.
\end{equation}
Thus, we hypothesize that magnetic field $B$ yields a crossover
between NFL and LFL regimes. It also reduces substantially the
relaxation rate as well as the thermal resistivity similar to
the case of normalized LMR of $\rm YbRh_2Si_2$, (see
Fig.~\ref{T12} B). Experimental data for LMR is taken from
\cite{gegmr}. Also of high priority are measurements of
low-energy inelastic neutron scattering on $\rm
ZnCu_3(OH)_6Cl_2$crystalline samples under the application of a
magnetic field, driving the system into the LFL sector of the
phase diagram (see Fig.~\ref{fig0}). The latter measurements
permit direct observation of a possible gap, since in this case
the impurity contribution is insignificant.

Recent neutron-scattering measurements on $\rm YbMgGaO_4$ reveal
broad spin excitations covering a wide region of the Brillouin
zone, thereby favoring the existence of a spinon Fermi surface
\cite{nat_qsl,prl_18}. At the same time, the measurements of
heat transport do not show any substantial magnetic excitations
contributions to thermal conductivity, thus raising doubt as to
presence of a QSL \cite{heat_tr}.  We speculate that the
observed behavior can be attributed to the emergence of a Mott
insulator (see e.g.\ \cite{prl_2016}), placing the system beyond
the FCQPT point in the phase diagram. At the same time, in the
case of herbertsmithite, the transport and thermodynamic
properties suggest that the gapless state represented by the
SCQSL is situated before FCQFT. We note that recent theoretical
studies of possible gaps in the ground state of herbertsmithite
lead to quite different conclusions about the presence of a gap
\cite{shaginyan:2011,science2011,prl2012,nat2012,Normand}.
Recent experimental studies also indicate that the spin-liquid
ground state in kagome lattice is gapless \cite{zorko}, and that
two distinct types of defects in herbertsmithite are found
\cite{prl2017}. This observation makes the impurity model
vulnerable. Therefore, to probe QSL in herbertsmithite reliably,
it is of crucial importance to carry out the experimental
studies suggested above.

\section{Charge transport and optical conductivity}

Our next step is analysis of the herbertsmithite optical
conductivity $\overline{\sigma}$ at low frequencies. To avoid
the contribution of phonon absorption to the conductivity, we
consider low temperatures $T$ and frequencies ${\omega}$
\cite{Pilon}. Under such assumptions, we can neglect the lattice
symmetry (kagome vs. triangular) as the wavelength is much
larger than the typical crystal size. In atomic units
$\hbar=c=1$, the Hamiltonian of a particle having momentum ${\bf
p} = i\nabla$, spin $\bf s$, and charge $e$ is given by
\begin{equation}\label{ham}
\hat{H}=\frac{1}{2m}\left({\bf p}-e{\bf
A}\right)^2+e\phi-\frac{\mu_B}{s}{\bf s}\cdot{\bf B},
\end{equation}
where ${\bf A}$ and ${\phi}$ are the vector and scalar
potentials respectively.  The nature of the vector potential
$\nabla{\bf A}=0$ implies that ${\bf p}$ and ${\bf A}$ operators
commute. Since for spinons $e=0$, only the last term on the
right-hand side of Eq.~\eqref{ham} contributes to
$\overline{\sigma}$.

Equations \eqref{MTT} and \eqref{SCHII} show that at small
$\omega$, the imaginary part of the spin susceptibility reads
\cite{shaginyan:2012:A}
\begin{equation}\label{chi2}
\chi{''}(\omega)\propto \omega(M^*)^2.
\end{equation}
Observing that the energy transfer $\varepsilon_B$
\cite{jltp:2017o,PinNoz} between magnetic field $B(\omega)$ and
our system is due to the last term in Eq.~\eqref{ham}, we obtain
\begin{equation}\label{dEDtB}
\varepsilon_B=2\pi\omega\left[\frac{\mu_B}{s}{\bf s}\cdot{\bf
B}\right]\chi{''}(\omega) \equiv 2\pi \frac{\omega^2
\mu_B^3}{s}(M^*)^2{\bf s}\cdot{\bf B}.
\end{equation}
The same quantity, $\varepsilon_E$, coming from the
electric field $E(\omega)$ reads
\begin{equation}\label{dEDtE}
\varepsilon_E=E^2(\omega)\overline{\sigma}(\omega).
\end{equation}
Comparison of Eqs.~\eqref{dEDtB} and \eqref{dEDtE} yields
\begin{equation}\label{sigma}
\overline{\sigma}(\omega)\propto\omega\chi{''}(\omega)\propto
\omega^2(M^*)^2.
\end{equation}
It follows from Eq.~\eqref{sigma} that
$\overline{\sigma}(\omega) \propto \omega^2$, and that behavior
is consistent with experimental facts obtained in measurements
on $\rm ZnCu_{3}(OH)_6Cl_2$ and $\rm EtMe_3Sb[Pd(dmit)_2]_2$
representing the best candidates for identification as materials
that host QSL \cite{Pilon,pustogow2018}. We note that $\chi''$
\eqref{chi2} at low $\omega$ coincides Eq.~\eqref{SCHII} and
provides a good description of the experimental data (see
Figs.~\ref{fig04} and \ref{fig05}). Equations \eqref{MTT} and
\eqref{sigma} inform us that at elevated temperatures
$\overline{\sigma}(T)$ is a decreasing function, which conforms
with the experimental data \cite{Pilon}. It also follows from
Eqs.~\eqref{MBB} and \eqref{sigma} that
$\overline{\sigma}(\omega,B)$ is a decreasing function of $B$.
This observation seems to be contradictory as no systematic $B$
dependence is observed experimentally \cite{Pilon}.

To elucidate the $\overline{\sigma}(B)$ behavior, we note that
corresponding experiments have been performed at $T$=3 K and
$B\leq 7$ T \cite{Pilon}.  This means that the system is still
in the transition region of the phase diagram and does not have
yet the normal Fermi liquid properties, where the effective mass
$M^*$ is given by Eq.~\eqref{MBB}. In this case the effective
mass properties is due to Eq.~\eqref{MTT}, rather than
Eq.~\eqref{MBB}. This means that the dependence
$\overline{\sigma}(B)$ cannot be observed. Accordingly, we think
that substantial  $\overline{\sigma}(B)$ dependence can be
observed at $B\simeq 7$ T in the case $T\leq 1$ K. In that case,
as one sees from Fig.~\ref{fig02}, the effective mass
$M^*\propto C_{\rm mag}/T$ is a diminishing function of the
applied magnetic field. Thus, we predict that
$\overline{\sigma}(B)$ diminishes at growing magnetic fields, as
follows from Eqs.~\eqref{MBB} and \eqref{sigma}. Since the
contribution coming from phonons does not depend on the magnetic
field, we propose that measurements of the variation
$\delta\sigma$, i.e.,
$\delta\sigma=\overline{\sigma}(\omega,B)-\overline{\sigma}(\omega,B=0)$,
can reveal both the physics of SCQSL and the ground state of
$\rm ZnCu_3(OH)_6Cl_2$, as well as the ground state of other
materials hosting a QSL. The above experiments on measurements
of the heat transport and optical conductivity can be carried
out on samples with varying $x$.  Such experiments yield
information on the influence of impurities on the value of the
gap. We predict that at moderate $x\sim 20$\% the SCQSL remains
robust, for both inhomogeneity and randomness facilitate
frustration.

\section{Summary}

The central message of this review is that to achieve a
satisfactory understanding of the quantum spin-liquid physics of
herbertsmithite at nanoscale, it is essential to perform the
targeted measurements on $\rm ZnCu_{3}(OH)_6Cl_2$ that we have
discussed. These focus on heat transport, low-energy inelastic
neutron scattering, and optical conductivity
$\overline{\sigma}$, in the presence of magnetic fields at low
temperatures. Moreover, we have suggested that the increasing
$x$, i.e. the percentage of $\rm Zn$ sites that are occupied by
$\rm Cu$, can facilitate the frustration of the lattice and
thereby act to stabilize the quantum spin liquid state. This
conjecture can be tested in experiments on samples of
herbertsmithite with different $x$ values under the application
of a magnetic field. To be specific, the results of the
measurements posed might yield an unambiguous answer to the
question whether the gap in spinon excitations really exists
and, if so, how it depends on impurities. Also, such experiments
will help to separate the universal effects from those coming
both from phonons and from the impurities that pollute each
specific sample of the material, and to represent frustrated
magnets with SCQSL as the new state of matter that exhibit the
properties of HF metals with one exception: These are
insulators, and do not support the charge current.

Other results reported in our review paper are based on the
description of low-frequency optical conductivity experimental
data in herbertsmithite. For that we implied that $\rm
ZnCu_{3}(OH)_6Cl_2$ is a system of strongly correlated fermions
with properties defined primarily by a quantum spin liquid of
chargeless spinons. We have also predicted the optical
conductivity dependence on magnetic field and pointed out the
explicit conditions under which such dependence can be
experimentally observed.  While making a step towards
confirmation of the SCQSL state existence in $\rm
ZnCu_{3}(OH)_6Cl_2$, our considerations may also pave the way to
studies of this state in other magnetic frustrated insulators at
nanoscale.

\section{Acknowledgements}
This work was partly supported by U.S. DOE, Division of Chemical
Sciences, Office of Basic Energy Sciences, Office of Energy
Research. JWC acknowledges support from the McDonnell Center for
the Space Sciences, and expresses gratitude to the University of
Madeira and its branch of Centro de Investiga\c{c}\~{a}o em
Matem\'{a}tica e Aplica\c{c}\~{o}es (CIMA) for gracious
hospitality during periods of extended residence.

\section{Declaration of interests}
The authors declare that they have no known competing financial
interests or personal relationships that could have appeared to
influence the work reported in this review paper.


\end{document}